# Approximating the least hypervolume contributor: NP-hard in general, but fast in practice

Karl Bringmann[*]     Tobias Friedrich[†]

**Abstract**

The hypervolume indicator is an increasingly popular set measure to compare the quality of two Pareto sets. The basic ingredient of most hypervolume indicator based optimization algorithms is the calculation of the hypervolume contribution of single solutions regarding a Pareto set. We show that exact calculation of the hypervolume contribution is **#P**-hard while its approximation is **NP**-hard. The same holds for the calculation of the minimal contribution. We also prove that it is **NP**-hard to decide whether a solution has the least hypervolume contribution. Even deciding whether the contribution of a solution is at most $(1+\varepsilon)$ times the minimal contribution is **NP**-hard. This implies that it is neither possible to efficiently find the least contributing solution (unless $\mathbf{P} = \mathbf{NP}$) nor to approximate it (unless $\mathbf{NP} = \mathbf{BPP}$).

Nevertheless, in the second part of the paper we present a fast approximation algorithm for this problem. We prove that for arbitrarily given $\varepsilon, \delta > 0$ it calculates a solution with contribution at most $(1 + \varepsilon)$ times the minimal contribution with probability at least $(1-\delta)$. Though it cannot run in polynomial time for all instances, it performs extremely fast on various benchmark datasets. The algorithm solves very large problem instances which are intractable for exact algorithms (e.g., 10000 solutions in 100 dimensions) within a few seconds.

## 1  Introduction

Multi-objective optimization deals with the task of optimizing several objective functions at the same time. As these functions are often conflicting, we cannot aim for a single optimal solution but for a set of Pareto optimal solutions. Unfortunately, the Pareto set frequently grows exponentially in the problem size. In this case, it is not possible to compute the whole front efficiently and the goal is to compute a good approximation of the Pareto front.

---

[*]Universität des Saarlandes, Saarbrücken, Germany
[†]Max-Planck-Institut für Informatik, Saarbrücken, Germany

There are many indicators to measure the quality of a Pareto set, but there is only one widely used that is strictly Pareto compliant [28], namely the hypervolume indicator. Strictly Pareto compliance means that given two Pareto sets $A$ and $B$ the indicator values $A$ higher than $B$ if the Pareto set $A$ dominates the Pareto set $B$. The hypervolume (HYP) measures the volume of the dominated portion of the objective space. It was first proposed and employed for multi-objective optimization by Zitzler and Thiele [26].

The hypervolume measure has become very popular recently and several algorithms have been developed to calculate it. The first one was the Hypervolume by Slicing Objectives (HSO) algorithm which was suggested independently by Zitzler [25] and Knowles [13]. To improve its runtime on practical instances, various speed up heuristics of HSO have been suggested [22, 24]. For $n$ points in $d$ dimensions, the currently best asymptotic runtime is $\mathcal{O}(n \log n + n^{d/2} \log n)$. It is obtained by Beume and Rudolph [2, 3] via an adaption of Overmars and Yap's algorithm [16] for Klee's Measure Problem [12]. There are also various algorithms for small dimensions [11, 15] and for calculating the contribution of a single point to the total hypervolume [5, 8].

From a geometric perspective, the hypervolume indicator is just measuring the volume of the union of a certain kind of boxes in $\mathbb{R}^d_{\geq 0}$, namely of boxes which share the reference point[1] as a common point. We will use the terms point and box interchangeably for solutions as the dominated volume of a point defines a box and vice versa. Given a set $M$ of $n$ points in $\mathbb{R}^d_{\geq 0}$, we define the hypervolume of $M$ to be

$$\mathrm{HYP}(M) := \mathrm{VOL}\left(\bigcup_{(x_1,\ldots,x_d)\in M} [0, x_1] \times \ldots \times [0, x_d]\right)$$

In [6, 7] the authors have proven that it is **#P**-hard[2] in the number of dimensions to calculate HYP precisely. Therefore, all hypervolume algorithms must have an exponential runtime in the number of objectives or boxes (unless $\mathbf{P} = \mathbf{NP}$). Without the widely accepted assumption $\mathbf{P} \neq \mathbf{NP}$, the only known lower bound for any $d$ is $\Omega(n \log n)$ [4]. Note that the worst-case combinatorial complexity (i.e., the number of faces of all dimensions on the boundary of the union) of $\Theta(n^d)$ does not imply any bounds on the computational complexity.

Though the **#P**-hardness of HYP dashes the hope for an exact subexponential algorithm, there are a few estimation algorithms [1, 6] for approximating the hypervolume based on Monte Carlo sampling. However, the only approximation algorithm with proven bounds is presented in [6]. There, the authors describe an FPRAS for HYP which gives an $\varepsilon$-approximation of the hypervolume with probability $(1 - \delta)$ in time $\mathcal{O}(\log(1/\delta)\, nd/\varepsilon^2)$.

---

[1] Without loss of generality we assume the reference point to be $0^d$.
[2] **#P** is the analog of **NP** for counting problems. For details see either the original paper by Valiant [21] or the standard textbook on computational complexity by Papadimitriou [17].

## New complexity results

We will now describe a few problems related to the calculation of the hypervolume indicator and state our results. For this, observe that calculating the hypervolume itself is actually not necessary in most hypervolume-based evolutionary multi-objective optimizers as for most algorithms it suffices to find a box with the minimal contribution to the hypervolume.

The *contribution* of a box $x \in M$ to the hypervolume of a set $M$ of boxes is the volume dominated by $x$ and no other element of $M$. We define the contribution $\mathrm{CON}(M, x)$ of $x$ to be

$$\mathrm{CON}(M, x) := \mathrm{HYP}(M) - \mathrm{HYP}(M \setminus x).$$

We are only aware of two algorithms which calculate $\mathrm{CON}(M, x)$ directly without the detour via $\mathrm{HYP}(M) - \mathrm{HYP}(M \setminus x)$ [5, 8]. In Section 2 we show that $\mathrm{CON}(M, x)$ is **#P**-hard to solve exactly. Furthermore, approximating CON by a factor of $2^{d^{1-\varepsilon}}$ is **NP**-hard for any $\varepsilon > 0$. Hence, CON is not approximable. Note that this is no contradiction to the above-mentioned FPRAS for HYP as an approximation of HYP does not yield an approximation of CON.

As a hypervolume-based optimizer is only interested in the box with the *minimal contribution*, we also consider the following problem. Given a set $M$ of $n$ boxes in $\mathbb{R}^d_{\geq 0}$, find the least contribution of any box in $M$, that is,

$$\mathrm{MINCON}(M) := \min_{x \in M} \mathrm{CON}(M, x).$$

The reduction in Section 2 shows that MINCON is **#P**-hard and not approximable, even if we know the box which is the least contributor.

Both mentioned problems can be used to find the box contributing the least hypervolume, but their hardness does not imply hardness of the problem itself, which we are trying to solve, namely calculating *which box has the least contribution*. Therefore we also examine the following problem. Given a set $M$ of $n$ boxes in $\mathbb{R}^d_{\geq 0}$, we want to find a box with the least contribution in $M$, that is,

$$\mathrm{LC}(M) := \operatorname*{argmin}_{x \in M} \mathrm{CON}(M, x).$$

If there are multiple boxes with the same (minimal) contribution, we are, of course, satisfied with any of them. In Section 2 we prove that this problem is **NP**-hard to decide, that is, for a given box one cannot decide whether it is the least contributor or not.

However, for practical purposes it most often suffices to solve a relaxed version of the above problem. That is, we just need to find a box which contributes not much more than the minimal contribution, meaning that it is only a $(1+\varepsilon)$ factor away. If we then throw out such a box, we have an error of at most $\varepsilon$. We will call this $\varepsilon$-$\mathrm{LC}(M)$ as it is an "approximation" of the problem LC. Given a set $M$ of $n$ boxes in $\mathbb{R}^d_{\geq 0}$ and $\varepsilon > 0$, we want to find a box with contribution at most $(1+\varepsilon)$ times the minimal contribution of any box in $M$, that is,

$$\mathrm{CON}(M, \varepsilon\text{-}\mathrm{LC}(M)) \leq (1+\varepsilon) \mathrm{MINCON}(M).$$

The final result of Section 2 is the **NP**-hardness of $\varepsilon$-LC. This shows, that there is no way of computing the least contributor efficiently, and even no way to approximate it.

### New approximation algorithm

In Section 3 we will give a "practical" algorithm for determining a small contributor. Technically speaking, it solves the following problem which we call $\varepsilon$-$\delta$-LC($M$): Given a set $M$ of $n$ boxes in $\mathbb{R}_{\geq 0}^d$, $\varepsilon > 0$ and $\delta > 0$, with probability at least $1 - \delta$ find a box with contribution at most $(1 + \varepsilon)\operatorname{MINCON}(M)$.

$$\Pr[\operatorname{CON}(M, \varepsilon\text{-}\delta\text{-LC}(M)) \leq (1 + \varepsilon)\operatorname{MINCON}(M)] \geq 1 - \delta.$$

As we will be able to choose $\delta$ arbitrarily, solving this problem is of high practical interest. By the **NP**-hardness of $\varepsilon$-LC there is no way of solving $\varepsilon$-$\delta$-LC efficiently, unless **NP** = **BPP**. This means, our algorithm cannot run in polynomial time for all instances. Its runtime depends on some hardness measure H (cf. Section 3.2), which is an intrinsic property of the given input, but generally unbounded, i.e., not bounded by some function in $n$ and $d$.

However, in Section 4 we show that our algorithm is practically very fast on various benchmark datasets, even for dimensions completely intractable for exact algorithms like $d = 100$ for which we can solve instances with $n = 10000$ points within seconds. This implies a huge shift in the practical usability of the hypervolume indicator.

## 2 Hardness of approximation

In this section we first show hardness of approximating MINCON, which we will use afterwards to show hardness of LC and $\varepsilon$-LC. We will reduce #MON-CNF to MINCON, which is the problem of counting the number of satisfying assignments of a Boolean formula in conjunctive normal form in which all variables are unnegated. While the problem of deciding satisfiability of such formula is trivial, counting the number of satisfying assignments is **#P**-hard and even approximating it by a factor of $2^{d^{1-\varepsilon}}$ for any $\varepsilon > 0$ is **NP**-hard, where $d$ is the number of variables (see Roth [20] for a proof).

**Theorem 1.** *MINCON is **#P**-hard and approximating it by a factor of $2^{d^{1-\varepsilon}}$ is **NP**-hard for any $\varepsilon > 0$.*

*Proof.* To show the theorem, we reduce #MON-CNF to MINCON. Let $\square(a_1, \ldots, a_d)$ denote a box $[0, a_1] \times \ldots \times [0, a_d]$. Let $f = \bigwedge_{k=1}^{n} \bigvee_{i \in C_k} x_i$ be a monotone Boolean formula given in CNF with $C_k \subseteq [d] := \{1, \ldots, d\}$, for $k \in [n]$, $d$ the number of variables, $n$ the number of clauses. First, we construct a box $A_k = \square(a_1^k, \ldots, a_d^k, 2^d + 2) \subseteq \mathbb{R}_{\geq 0}^{d+1}$ for each clause $C_k$ with one vertex at the origin and the opposite vertex at $\overline{(a_1^k, \ldots, a_d^k, 2^d + 2)}$, where we set

$$a_i^k = \begin{cases} 1, & \text{if } i \in C_k \\ 2, & \text{otherwise} \end{cases}, \quad i \in [d].$$

Additionally, we need a box $B = \Box(2, \ldots, 2, 1) \subseteq \mathbb{R}^{d+1}_{\geq 0}$ and the set $M = \{A_1, \ldots, A_n, B\}$. Since we can assume without loss of generality that no clause is dominated by another, meaning $C_i \not\subseteq C_j$ for every $i \neq j$, every box $A_k$ overlaps uniquely a region $[x_1, x_1 + 1] \times \ldots \times [x_d, x_d + 1] \times [1, 2^d + 2]$ with $x_i \in \{0, 1\}$, $i \in [d]$, so that the contribution of every box $A_k$ is greater than $2^d$ and the contribution of $B$ is at most $2^d$, so that $B$ is indeed the least contributor.

Observe that the contribution of $B$ to $\mathrm{HYP}(M)$ can be written as a union of boxes of the form $B_x = [x_1, x_1 + 1] \times \cdots \times [x_d, x_d + 1] \times [0, 1]$ with $x = (x_1, \ldots, x_d) \in \{0, 1\}^d$. Let $x \in \{0, 1\}^d$. We will now show that $B_x$ is a subset of the contribution of $B$ to $\mathrm{HYP}(M)$ if and only if $x$ satisfies $f$.

First, assume that $B_x$ is not a subset of the contribution of $B$ to $\mathrm{HYP}(M)$. Then it is a subset of $\bigcup_{k=1}^n A_k$ and hence a subset of some $A_k$. This implies that $a_i^k \geq x_i + 1$ for all $i \in [d]$ and $i \notin C_k$ for all $i$ with $x_i = 1$. In other words, $x$ satisfies $\bigwedge_{i \in C_k} \neg x_i$ for some $k$. This proves then that $x$ satisfies the negated formula $\bar{f} = \bigvee_{k=1}^n \bigwedge_{i \in C_k} \neg x_i$.

The same holds in the opposite direction, that is, if $B_x$ is a subset of the contribution of $B$, then $x$ satisfies $f$. Together with the fact $\mathrm{VOL}(B_x) = 1$ this yields $\mathrm{MINCON}(M) = \mathrm{CON}(M, B) = |\{x \in \{0, 1\}^d \mid x \text{ satisfies } f\}|$ which implies a polynomial time algorithm solving $\mathrm{MINCON}(M)$. This finishes the proof as this would result in a polynomial time algorithm for #MON-CNF. □

Note that the reduction from above implies that MINCON is **#P**-hard and **NP**-hard to approximate even if the least contributor is known. Moreover, since we constructed boxes with integer coordinates in $[0, 2^d + 2]$ a number of $b = \mathcal{O}(d^2 n)$ bits suffices to represent all $d+1$ coordinates of the $n+1$ constructed points. Hence, MINCON is hard even if all coordinates are integral. We define as input size $b + n + d$, where $b$ is the number of bits in the input. We will use this result in the next proof. Also note that the same hardness for CON follows immediately, as it is hard to compute $\mathrm{CON}(M, B)$ as constructed above.

By reducing MINCON to LC, one can now show **NP**-hardness of LC. We skip this proof and directly prove **NP**-hardness of $\varepsilon$-LC by using the hardness of approximating MINCON in the following theorem.

**Theorem 2.** $\varepsilon$-LC *is* **NP**-*hard for any constant* $\varepsilon$. *More precisely, it is* **NP**-*hard for* $(1 + \varepsilon)$ *bounded from above by* $2^{d^{1-c} - 1}$ *for some* $c > 0$.

*Proof.* We reduce MINCON to $\varepsilon$-LC. Let $M$ be a set of $n$ boxes in $\mathbb{R}^d_{\geq 0}$, i.e., a problem instance of MINCON represented by a number of $b$ bits, so that the input size is $b + n + d$.

As discussed above, we can assume that the coordinates are integral. We can further assume that $d \geq 2$ as MINCON is trivial for $d = 1$. The minimal contribution of $M$ might be 0, but this occurs if and only if one box in $M$ dominates another. As the latter can be checked in polynomial time, we can without loss of generality also assume that $\mathrm{MINCON}(M) > 0$.

Now, let $V$ be the volume of the bounding box of all the boxes in $M$, i.e., the product of all maximal coordinates in the $d$ dimensions. We know that $V$ is an integer with $1 \leq V \leq 2^b$, as there are only $b$ bits in the input.

We now define a slightly modified set of boxes:

$$\begin{aligned}
A &= \{\Box(a_1 + 2V, a_2, \ldots, a_d) \mid \Box(a_1, \ldots, a_d) \in M\}, \\
B &= \Box(2V, \ldots, 2V), \\
C_\lambda &= \Box(1, \ldots, 1, 2V + \lambda), \\
M_\lambda &= A \cup \{B\} \cup \{C_\lambda\}.
\end{aligned}$$

The boxes in $A$ are the boxes of $M$, but shifted along the $x_1$-axis. By definition, $a_i \leqslant V$, $i \in [d]$ for all $\Box(a_1, \ldots, a_d) \in M$. The contribution to $\mathrm{HYP}(M_\lambda)$ of a box in $A$ is the same as the contribution to $\mathrm{HYP}(M)$ of the corresponding box in $M$ as the additional part is overlapped by the "blocking" box $B$. Also note that the contribution of a box in $A$ is less than or equal to $V$.

The box $B$ uniquely overlaps at least the space $[V, 2V] \times \ldots \times [V, 2V]$ (as every coordinate of a point in $M$ is less than equal to $V$) which has volume at least $V$. Hence, $B$ is never the least contributor of $M_\lambda$. The box $C_\lambda$ then has a contribution of $\mathrm{VOL}([0,1] \times \ldots \times [0,1] \times [2V, 2V + \lambda]) = \lambda$, so that $C_\lambda$ is a least contributor if and only if $\lambda$ is less than or equal to the minimal contribution of any box in $A$ to $\mathrm{HYP}(M_\lambda)$ which holds if and only if we have $\lambda \leq \mathrm{MINCON}(M)$.

As we can decide whether $C_\lambda$ is the least contributor by one call to $\mathrm{LC}(M_\lambda)$, we can do a sort of a binary search on $\lambda$. As we are interested in a multiplicative approximation, we search for $\kappa := \log_2(\lambda)$ to be the largest value less than equal to $\log_2(\mathrm{MINCON}(M))$, where $\kappa$ now is an integer in the range $[0, b]$. Since we can only answer $\varepsilon$-LC-queries, we cannot do exact binary search. However, we can still follow its lines, recurring on the left half of the current interval, if for the median value $\kappa_\mathrm{m}$ we get $\varepsilon\text{-LC}(M_{\lambda_\mathrm{m}}) = C_{\lambda_\mathrm{m}}$, where $\lambda_\mathrm{m} = 2^{\kappa_\mathrm{m}}$, and on the right half, if we get any other result.

The incorrectness of $\varepsilon$-LC may misguide our search, but since we have

$$\mathrm{CON}(M, \varepsilon\text{-LC}(M)) \leq (1 + \varepsilon)\,\mathrm{MINCON}(M)$$

it can give a wrong answer (i.e., not the least contributor) only if we have $(1+\varepsilon)^{-1}\mathrm{MINCON}(M) \leq 2^\kappa \leq (1+\varepsilon)\,\mathrm{MINCON}(M)$. Outside of this interval our search goes perfectly well. Thus, after the binary search, i.e, after at most $\lceil \log_2(b) \rceil$ many calls to $\varepsilon$-LC, we end up at a value $\kappa$ which is either inside the above interval (in which case we are satisfied) or the largest integer smaller than $\log_2((1+\varepsilon)^{-1}\mathrm{MINCON}(M))$ or the smallest integer greater than $\log_2((1+\varepsilon)\,\mathrm{MINCON}(M))$. Hence, we have

$$\kappa \leq \log_2((1+\varepsilon)\,\mathrm{MINCON}(M)) + 1$$

implying

$$\lambda = 2^\kappa \leq 2(1+\varepsilon)\,\mathrm{MINCON}(M).$$

Analogously, we get

$$\lambda = 2^\kappa \geq \frac{1}{2(1+\varepsilon)}\mathrm{MINCON}(M).$$

Therefore after $\mathcal{O}(\log(b))$ many calls to $\varepsilon$-LC we get a $2\,(1+\varepsilon)$ approximation of MINCON($M$). Since this is **NP**-hard for $2\,(1+\varepsilon)$ bounded from above by $2^{d^{1-c}}$ for some $c > 0$, we showed **NP**-hardness of $\varepsilon$-LC in this case. Note that this includes any constant $\varepsilon$. □

The **NP**-hardness of $\varepsilon$-LC not only implies **NP**-hardness of LC, but also the non-existence of an efficient algorithm for $\varepsilon$-$\delta$-LC unless **NP** = **BPP**. The above proof also gives a very good intuition about the problem $\varepsilon$-LC: As we can approximate the minimal contribution by a small number of calls to $\varepsilon$-LC, there cannot be a much faster way to solve $\varepsilon$-LC but to approximate the contributions – approximating at least the least contribution can be only a factor of $\mathcal{O}(\log(b))$ slower than solving $\varepsilon$-LC. This motivates the algorithm we present in the next section, which tries to approximate the contributions of the various boxes.

## 3 Practical approximation algorithm

The last section ruled out the possibility of a worst case efficient algorithm for computing or approximating the least contributor. Nevertheless, we are now presenting an algorithm $\mathcal{A}$ that is "safe" and has a good practical runtime, but no polynomial worst case runtime (as this is not possible). By "safe" we mean that it provably solves $\varepsilon$-$\delta$-LC, i.e., it holds that

$$\Pr[\text{CON}(M, \mathcal{A}(M, \varepsilon, \delta)) \leq (1+\varepsilon)\,\text{MINCON}(M)] \geq 1 - \delta. \qquad (1)$$

We consider an $\varepsilon$ around $10^{-2}$ or $10^{-3}$ as sufficient for typical instances. This implies for most instances that we return the correct result as there are no two small contributions which are only a $(1+\varepsilon)$-factor apart. For the remaining cases we return at least a box which has contribution at most $(1+\varepsilon)$ times the minimal contribution, which means we make an "error" of $\varepsilon$.

Additionally, the algorithm is going to be a randomized Monte Carlo algorithm, which is why we need the failure probability $\delta$ and do not always return the correct result. However, we will be able to set $\delta = 10^{-6}$ or even $\delta = 10^{-12}$ without increasing the runtime overly. In the following we will describe algorithm $\mathcal{A}$, prove its correctness and describe its runtime.

### 3.1 The algorithm $\mathcal{A}$

Our algorithm works as follows. First, it is essential to determine for each box $A$ the minimal bounding box of the space that is uniquely overlapped by the box. To do so we start with the box $A$ itself. Then we iterate over all other boxes $B$. If $B$ dominates $A$ in all but one dimension, then we can cut the bounding box in the non-dominated dimension. This can be realized in time $\mathcal{O}(d\,n^2)$.

Having the bounding box $\text{BB}_A$ of the contribution of $A$ we start to sample randomly in it. For each random point we determine if it is uniquely dominated by $A$. If we checked NOSAMPLES($A$) random points and NOSUCCSAMPLES($A$) of them were uniquely dominated by $A$, then the contribution of $A$ is about

$$\widetilde{V}_A := \frac{\text{NOSUCCSAMPLES}(A)}{\text{NOSAMPLES}(A)} \text{VOL}(BB_A), \qquad (2)$$

where $\text{VOL}(BB_A)$ denotes the volume of the bounding box of the contribution of $A$. Additionally, we can give an estimate of the deviation of $\widetilde{V}_A$ from $V_A$, the correct contribution of $A$ (i.e., $V_A = \text{CON}(M, A)$): Using Chernoff's inequality we get that for

$$\Delta_R(A) := \sqrt{\frac{\log(2nR^{1+\gamma}\delta^{-1}(1+\gamma)/\gamma)}{2\text{NOSAMPLES}(A)}} \text{VOL}(BB_A) \qquad (3)$$

the probability that $V_A$ deviates from $\widetilde{V}_A$ by more than $\Delta_R(A)$ is small enough. Here, as usual $n$ is the number od boxes and $\delta$ is the probability of error we want to have overall. Additionally, $R$ is a variable parameter (the round we are in), and $\gamma \in (0, 1]$ is a constant, which one can adjust to get the best performance. The log-factor is chosen such that the analysis of the algorithm works out. Note that independently a similar sampling approach has been described in another context in [19].

We would like to sample in the bounding boxes in parallel such that every $\widetilde{V}_A$ deviates about the same $\Delta$. We do this in rounds: In the first round we initialize $\Delta = \Delta_1$ arbitrarily (e.g., $\Delta_1 = \max_{A \in M} \text{VOL}(BB_A)$). In every other round $R$ we decrease $\Delta$ by a constant factor, e.g., $\Delta_R = \frac{1}{2}\Delta_{R-1}$. Then we sample in each bounding box until we have $\Delta_R(A) \leq \Delta_R$ for each box $A$. If we then have at any point two boxes $A$ and $B$ with

$$\widetilde{V}_A - \Delta_R(A) > \widetilde{V}_B + \Delta_R(B) \qquad (4)$$

we can with good probability assume that $A$ is not a least contributor, as we would need to have $\widetilde{V}_A - V_A > \Delta_R(A)$ or $V_B - \widetilde{V}_B > \Delta_R(B)$ for $A$ having a smaller contribution than $B$ (which is necessary for $A$ being the least contributor). Hence, in such cases we can delete $A$ from our race, meaning that we do not have to sample in its bounding box anymore. Note that we never have to compare two arbitrary boxes, but only a box $A$ to the currently smallest box $\widetilde{LC}$, i.e., the box with $\widetilde{V}_{\widetilde{LC}}$ minimal.

We can run this race, deleting boxes if their contribution is clearly too much by the above selection equation until either there is just one box left, in which case we have found the least contributor, or until we have reached a point where we have approximated all contributions well enough. Given an abortion criterion $\varepsilon$ we can just return $\widetilde{LC}$ (the box with currently smallest approximated contribution) when we have (being in round $R$)

$$\widetilde{V}_A - \Delta_R(A) > 0 \text{ and } \frac{\widetilde{V}_{\widetilde{LC}} + \Delta_R(\widetilde{LC})}{\widetilde{V}_A - \Delta_R(A)} \leq 1 + \varepsilon,$$

for any box $A \neq \widetilde{LC}$ still in the race. If this equation holds, then we can be quite sure that any box has contribution at least $\frac{1}{1+\varepsilon}V_{\widetilde{LC}}$. So, returning $\widetilde{LC}$, we have solved $\varepsilon$-$\delta$-LC after all.

**Algorithm 1** $\mathcal{A}(M, \varepsilon, \delta)$ solves $\varepsilon$-$\delta$-LC$(M)$ for a set $M$ of $n$ boxes in $\mathbb{R}^d_{\geq 0}$ and $\varepsilon, \delta > 0$, i.e., it determines a box $x \in M$ such that $\Pr[\text{CON}(M, x) \leq (1 + \varepsilon) \text{MINCON}(M)] \geq 1 - \delta$.

> determine the bounding boxes $\text{BB}_A$ for all $A \in M$
> initialize NOSAMPLES$(A)$ = NOSUCCSAMPLES$(A)$ = 0 for all $A \in M$
> initialize $R = 0, \Delta_R = \max_{A \in M} \text{VOL}(BB_A)$
> set $S := M$
> **repeat**
>   set $\Delta_{R+1} := \Delta_R / 2$
>   set $R := R + 1$
>   **for all** $A \in S$ **do**
>     **repeat**
>       sample a random point in $\text{BB}_A$
>       increase NOSAMPLES$(A)$ and possibly NOSUCCSAMPLES$(A)$
>       update $\widetilde{V}_A$ and $\Delta_R(A)$ according to (2) and (3)
>     **until** $\Delta_R(A) \leq \Delta_R$
>   **od**
>   set $\widetilde{LC} := \arg\min\{\widetilde{V}_A \mid A \in S\}$
>   assume $|\widetilde{V}_A - V_A| \leq \Delta_R(A)$ for all $A \in S$
>   **for all** $A \in S$ **do**
>     **if** $\widetilde{V}_A - \Delta_R(A) > \widetilde{V}_{\widetilde{LC}} + \Delta_R(\widetilde{LC})$ **then**
>       $S := S \setminus \{A\}$
>   **od**
> **od**
> **until** $|S| = 1$ or $(\widetilde{V}_A - \Delta_R(A) > 0$ and $\frac{\widetilde{V}_{\widetilde{LC}} + \Delta_R(\widetilde{LC})}{\widetilde{V}_A - \Delta_R(A)} \leq 1 + \varepsilon$ $\forall \widetilde{LC} \neq A \in S)$
> **return** $\widetilde{LC}$

Above you can find in pseudo code what we just described. The only new thing is a line indicating that the algorithm makes an assumption about the $\widetilde{V}_A$'s. This line will help in the proof of correctness: First, it can easily be seen that the algorithm solves $\varepsilon$-LC if all assumptions being made are true. In a second step we have to bound the probability of any assumption to be wrong, to show that the algorithm solves $\varepsilon$-$\delta$-LC.

### 3.2 Runtime

As discussed above, our algorithm needs a runtime of at least $\Omega(dn^2)$. This seems to be the true runtime on many practical instances (cf. Section 4). However, by Theorem 2 we cannot hope for a matching upper bound. In this section we present an upper bound on the runtime depending on some characteristics of the input.

For an upper bound, observe that we have to approximate each box $A$ up to $\Delta = \mathcal{O}(V_A - \text{MINCON}(M))$ to be able to delete it. One can also show that the expected value of $\Delta$ where we delete box $A$ is $\Omega(V_A - \text{MINCON}(M))$. By equation (3) solved for NOSAMPLES($A$) we observe that we need a number of

$$\frac{\log(2nR^{1+\gamma}\delta^{-1}(1+\gamma)/\gamma)\text{VOL}(\text{BB}_A)^2}{\Omega(V_A - \text{MINCON}(M))^2} = \mathcal{O}\left(\frac{\log(nR^{1+\gamma}\delta^{-1})\text{VOL}(\text{BB}_A)^2}{(V_A - \text{MINCON}(M))^2}\right)$$

samples to delete box $A$ on average. For the least contributor $LC$, we need $\mathcal{O}\left(\frac{\log(nR^{1+\gamma}\delta^{-1})\text{VOL}(\text{BB}_{LC})^2}{(\text{sec-min}(V) - \text{MINCON}(M))^2}\right)$ many samples until we have finally deleted all other boxes, where sec-min($V$) denotes the second smallest contribution of any box in $M$. We have $\Delta_R = 2^{-R} \max_{A \in M} \text{VOL}(BB_A)$ so that $\Delta = \mathcal{O}(V_A - \text{MINCON}(M))$ happens for $R = \Theta\left(\log\left(\frac{\max_{A \in M} \text{VOL}(BB_A)}{V_A - \text{MINCON}(M)}\right)\right)$. Putting this into above bounds we get an upper bound for the expected number of samples of our algorithm. Since each sample takes runtime $\mathcal{O}(dn)$ and everything besides the sampling takes much less runtime, we get an overall runtime of

$$\mathcal{O}(dn\,(n + \text{H})),$$

where

$$\begin{aligned}
\text{H} :=& \frac{\text{VOL}(\text{BB}_{LC})^2}{(\text{sec-min}(V) - \text{MINCON}(M))^2} \\
& \cdot \left(\log(n/\delta) + \log\log\left(\frac{\max_{A \in M} \text{VOL}(BB_A)}{\text{sec-min}(V) - \text{MINCON}(M)}\right)\right) \\
& + \sum_{LC \neq A \in S} \frac{\text{VOL}(\text{BB}_A)^2}{(V_A - \text{MINCON}(M))^2} \\
& \cdot \left(\log(n/\delta) + \log\log\left(\frac{\max_{A \in M} \text{VOL}(BB_A)}{V_A - \text{MINCON}(M)}\right)\right)
\end{aligned}$$

is a certain measure of hardness of the input. As we have $\text{VOL}(\text{BB}_A) \geq V_A$ we conclude that $\text{H} \geq (n-1)\log(n/\delta)$. Of course, the log log-factors do not increase the hardness too much. Focussing on the first factors, the hardness H is small, if we have for all boxes $\text{VOL}(\text{BB}_A) \approx V_A$ and $\text{MINCON}(M) \ll V_A$. On the other hand, this value is large if we are in one of the following two situations: First, there may be a point with a large bounding box $\text{VOL}(\text{BB}_A)$ but a small contribution $V_A$. Cases where the ratio of the two is arbitrarily large can easily be constructed. Second, there may be two or more boxes contributing the minimal contribution or only slightly more than it. In this case the value $V_A - \text{MINCON}(M)$ is small. These two situations are the hard cases for our algorithm. However, we observed empirically that in random instances these cases rarely occur.

To be precise, the hardness H may even be undefined: If there are two minimal contributors, then $V_A - \text{MINCON}(M) = 0$ for one of the two boxes, so that we divide by 0. This clearly has to be the case, as we can never decide of two contributions whether they are equal or just nearly equal, if the difference is tiny. In this case our abortion criterion comes into play: With high enough probability after approximating every contribution up to $\Delta = \frac{\varepsilon}{4+2\varepsilon}\text{MINCON}(M)$ we have $\widetilde{V}_{LC} \leq V_{LC} + \Delta$, thus $\widetilde{V}_{\widetilde{LC}} \leq V_{LC} + \Delta$, and $\widetilde{V}_A \geq V_{LC} - \Delta$ for every other box $A$ still in the race. Then we conclude

$$\frac{\widetilde{V}_{\widetilde{LC}} + \Delta(\widetilde{LC})}{\widetilde{V}_A - \Delta(A)} \leq \frac{V_{LC} + 2\Delta}{V_{LC} - 2\Delta} = \frac{1 + 2\frac{\varepsilon}{4+2\varepsilon}}{1 - 2\frac{\varepsilon}{4+2\varepsilon}} = 1 + \varepsilon$$

for every box $\widetilde{LC} \neq A \in S$, so that we return a $(1 + \varepsilon)$-approximation. Hence, the above defined value for $\Delta$ suffices to enforce abortion. As we get this $\Delta$ after $\text{NOSAMPLES}(A) = \frac{\log(2nR^{1+\gamma}\delta^{-1}(1+\gamma)/\gamma)\text{VOL}(BB_A)^2}{2(\frac{\varepsilon}{4+2\varepsilon}\text{MINCON}(M))^2}$ samples and $R = \log\left(\frac{(4+2\varepsilon)\max_{A \in M}\text{VOL}(BB_A)}{\varepsilon \cdot \text{MINCON}(M)}\right)$, this yields another upper bound for the overall number of samples, a still unbounded but always finite value:

$$\mathcal{O}\Bigg(\Big(\log(n/\delta) + \log\log\Big(\frac{\max_{A \in M}\text{VOL}(BB_A)}{\varepsilon \cdot \text{MINCON}(M)}\Big)\Big) \sum_{A \in M} \frac{\text{VOL}(BB_A)^2}{\varepsilon^2 \text{MINCON}(M)^2}\Bigg)$$

However, for the random testcases that we consider in Section 4 the above defined hardness H is a more realistic measure of runtime as there are never two identical contributions, not too many equally small contributions and the bounding box is never too much larger than the contribution. There one observes values for H that roughly lie in the interval $[n\log(n/\delta), 10n\log(n/\delta)]$.

### 3.3  Correctness of our algorithm

To prove correctness of our algorithm we need to show two things: First, if all the assumptions the algorithm makes (see the line "assume $|\widetilde{V}_A - V_A| \leq \Delta_R(A)$ for all $A \in S$") turn out to be true, then it returns a box $X$ contributing not more than $(1 + \varepsilon)\text{MINCON}(M)$, meaning that it solves $\varepsilon$-LC. Second, the probability of any assumption being wrong is small. The next two lemmas will show these two statements:

**Lemma 3.** *If all assumptions made by the algorithm are true and the algorithm terminates, then it solves $\varepsilon$-LC.*

*Proof.* Assume the assumptions $|\widetilde{V}_A - V_A| \leq \Delta_R(A)$ and $|\widetilde{V}_B - V_B| \leq \Delta_R(B)$ are true. At the point at which the selection criterion is met, i.e., we have $\widetilde{V}_A - \Delta_R(A) > \widetilde{V}_B + \Delta_R(B)$, we conclude that

$$V_A \geq \widetilde{V}_A - \Delta_R(A) > \widetilde{V}_B + \Delta_R(B) \geq V_B,$$

so that we have $V_A > V_B$. Hence, the box $A$ clearly cannot be a least contributor and we can discard it from our race.

Furthermore, suppose that we have $\widetilde{V}_A - \Delta_R(A) > 0$ and $\frac{\widetilde{V}_B + \Delta_R(B)}{\widetilde{V}_A - \Delta_R(A)} \leq 1 + \varepsilon$. We can conclude that

$$\frac{V_B}{V_A} \leq \frac{\widetilde{V}_B + \Delta_R(B)}{\widetilde{V}_A - \Delta_R(A)} \leq 1 + \varepsilon,$$

meaning that we have $V_B \leq (1+\varepsilon)V_A$. Using this for $B = \widetilde{LC}$ we get that $\widetilde{LC}$ really is a box contributing not more than $(1+\varepsilon)\mathrm{MINCON}(M)$ if the abortion criterion is met. This shows correctness of the algorithm, as long as all assumptions being made are true. □

We now show that the probability of any assumption of the algorithm being wrong is small:

**Lemma 4.** *The probability of any assumption made by the algorithm being wrong is at most $\delta$.*

*Proof.* For any box $A$ in every round $R$ we make an assumption $\mathbb{A}(A,R)$ on $\widetilde{V}_A$, given that $A$ survived until round $R$. Observe that this assumption is made when $\mathrm{NOSAMPLES}(A) = m_{A,R}$ for some deterministically determined $m_{A,R}$ (meaning that it is no random variable). Hence, we can bound as follows:

$$\Pr[\text{assumption } \mathbb{A}(A,R) \text{ is wrong}]$$
$$= \Pr\left[A \text{ survives until round } R \text{ and } |\widetilde{V}_A - V_A| > \Delta_R(A) \mid \mathrm{NOSAMPLES}(A) = m_{A,R}\right]$$
$$\leq \Pr[|\widetilde{V}_A - V_A| > \Delta_R(A) \mid \mathrm{NOSAMPLES}(A) = m_{A,R}].$$

We use the definition of $\widetilde{V}_A = \frac{\mathrm{NOSUCCSAMPLES}(A)}{\mathrm{NOSAMPLES}(A)} \mathrm{VOL}(BB_A)$ and write $\mathrm{NOSUCCSAMPLES}(A)$ as a sum of independent identically distributed random variables $X_i$ with $X_i = 1$ indicating that the $i$-th sample was successful and $X_i = 0$ otherwise. Putting in the definition of $\Delta_R(A)$ we get

$$= \Pr\left[\left|\sum_{i=1}^{m_{A,R}} X_i \cdot \frac{\mathrm{VOL}(BB_A)}{m_{A,R}} - V_A\right| > \sqrt{\frac{\log(2nR^{1+\gamma}\delta^{-1}(1+\gamma)/\gamma)}{2m_{A,R}}} \mathrm{VOL}(BB_A)\right]$$
$$= \Pr\left[\left|\sum_{i=1}^{m_{A,R}} X_i - m_{A,R}\frac{V_A}{\mathrm{VOL}(BB_A)}\right| > \sqrt{\log(2nR^{1+\gamma}\delta^{-1}(1+\gamma)/\gamma)m_{A,R}/2}\right].$$

We observe that $\frac{V_A}{\mathrm{VOL}(BB_A)}$ is the expected value of each $X_i$. This allows us to use Chernoff's inequality which states that we have for $X = \sum_{i=1}^{m_{A,R}} X_i$ the inequality $\Pr[|X - \mathrm{E}[X]| > a] \leq 2\exp(-2a^2/m_{A,R})$. In our case this yields

$$\Pr[\text{assumption } \mathbb{A}(A,R) \text{ is wrong}] \leq \frac{\gamma\delta}{nR^{1+\gamma}(1+\gamma)}.$$

We use the Union Bound to bound the probability of any assumption being

wrong:

$$\Pr[\text{any assumption is wrong}]$$
$$\leq \sum_{A \in M} \sum_{R=1}^{\infty} \Pr[\text{assumption } \mathbb{A}(A,R) \text{ is wrong}]$$
$$\leq \sum_{A \in M} \sum_{R=1}^{\infty} \frac{\gamma \delta}{n R^{1+\gamma}(1+\gamma)}$$
$$= \sum_{R=1}^{\infty} \frac{\gamma \delta}{R^{1+\gamma}(1+\gamma)},$$

as $|M| = n$. One can easily bound

$$\sum_{R=1}^{\infty} \frac{1}{R^{1+\gamma}} \leq 1 + \int_{1}^{\infty} \frac{1}{x^{1+\gamma}} dx = 1 + 1/\gamma.$$

Using this, we finally get

$$\Pr[\text{any assumption is wrong}] \leq \delta. \qquad \square$$

Both lemmas together directly imply the correctness of our algorithm.

**Corollary 5** (Correctness of $\mathcal{A}$). *The probability of $\mathcal{A}(M,\varepsilon,\delta)$ being a correct result of $\varepsilon$-LC is at least $(1-\delta)$, i.e.,*

$$\Pr[\text{CON}(M, \mathcal{A}(M,\varepsilon,\delta)) \leq (1+\varepsilon) \text{MINCON}(M)] \geq 1 - \delta.$$

### 3.4 Heuristical improvements

To increase the practical efficiency of our algorithm, we implemented a few further optimizations that decrease the actual runtime. In this section we will describe three implemented heuristics.

#### 3.4.1 Push on $\Delta(\widetilde{LC})$:

Since we compare all boxes still in the race with the currently minimal one $\widetilde{LC}$, it is intuitively a good idea to decrease $\Delta(\widetilde{LC})$ faster than all other $\Delta(A)$, i.e., if we have a current bound of $\Delta(A) \leq \Delta$ for any $A \in S$ we should sample in $\widetilde{LC}$ until we have $\Delta(\widetilde{LC}) \leq \alpha \Delta$ for some constant $\alpha < 1$. This improves the runtime by up to a factor of 4: If we needed some value of $\Delta$ to distinguish boxes $A$ and $\widetilde{LC}$ before, we now only need $\Delta' = \frac{2}{1+\alpha}\Delta$ for $A$ and $\Delta'(\widetilde{LC}) = \frac{2\alpha}{1+\alpha}\Delta$. As the number of samples needed is proportional to $\Delta^{-2}$ it changes by a good factor of $\frac{1+\alpha^2}{4} \approx \frac{1}{4}$ for $n-1$ boxes and a worse factor of $\frac{1+\alpha^2}{4\alpha^2}$ for the one box $\widetilde{LC}$. On practical instances $\alpha = 0.2$ seemed to be a reasonable value.

#### 3.4.2 Sampling heuristic:

It is clear how to find a random point $X$ inside a bounding box $\mathrm{BB}_A$. Then we have to check whether $X$ lies in a box $A \neq B \in M$. If no $B$ dominates $X$, then $X$ is counted as a successful sample, otherwise not. Now it suffices to test whether $X$ lies in a subset of $M$. Only points with all coordinates bigger than the lower vertex of the bounding box $\mathrm{BB}_A$ can possibly dominate $X$. By determining these once at the beginning and saving them we get a space requirement of $\mathcal{O}(n^2)$ but an improvement of the runtime of between one and two orders of magnitude. Furthermore, we can decide to rearrange these points such that we check whether $X$ lies in all possible dominating boxes $B$ in descending order of the volume of the part of $\mathrm{BB}_A$ that is dominated by $B$. This way we intuitively speed up all "unsuccessful" searches, i.e., all samples where $X$ is indeed dominated. On real instances this yields another speedup of small constant factor.

#### 3.4.3 Exact calculation:

As an involved sampling algorithm only makes sense for large instances, our implementation uses a classical exact algorithm for small $n$ and $d$. The difficulty is to decide when to do so. Our approach works as follows. After we determined the boxes that dominate the lower vertex of a bounding box $\mathrm{BB}_A$, i.e., the boxes that "influence" the contribution of $A$, some of those sets of influencing boxes have quite a small cardinality $n_A$. Especially boxes with small contribution tend to have only a small number of influencing boxes. Hence if this number is small we can determine the contribution exactly by using some classical hypervolume algorithm. In this case, we just restrict the $n_A$ influencing boxes to the bounding box $\mathrm{BB}_A$ and solve the induced HYP problem (inside $\mathrm{BB}_A$) to get a volume $v$. After that, we subtract $v$ from $\mathrm{VOL}(\mathrm{BB}_A)$ which gives us the correct contribution of $A$.

This can be done for any box $A$ with a small number of influencing boxes $n_A$. After calculating its contribution exactly we just have to set $\Delta(A) := 0$ which works fine in our algorithm. The only problem is to decide which values of $n_A$ are to be considered "small" in this respect. For each box $A$ we count how many elementary operations we made so far for sampling in its bounding box (by counting how many coordinate comparisons we made), calling this number $\mathrm{NOOPS}(A)$. We also try to estimate the runtime (number of elementary operations) we would need for computing the contribution of $A$ exactly. This, of course, depends on the algorithm one uses. We use the well-known HSO algorithm by Zitzler [25] and the algorithm by Beume and Rudolph [2, 3] which we will call BR, but one can, of course, use an arbitrary exact hypervolume algorithm. For those algorithms we can bound the runtime by $\mathcal{O}(n\binom{n+d-2}{d-1})$ for HSO [23] and $\mathcal{O}(n \log n + n^{d/2} \log n)$ for BR [2, 3]. By estimating the constant hidden in the asymptotic notation, we get an upper bound of the number of operations the two algorithms make. Having this, we can at any point in the algorithm decide to compute a contribution exactly rather than continue to sample in it. That is, if $\text{NOOPS}(A) > \text{estimatedRuntimeHSO}(n_A, d)$ we just compute it exactly. This way we need only twice as much time as if we had computed the contribution exactly right from the start. Also, if we needed only a small number of samples more to throw $A$ out of the race, we only needed twice as much time overall computing the contribution exactly than continuing to sample. Hence, by this decision we always need at most twice the number of operations we would have needed with the optimal decision. This also implies that asymptotically our runtime with this heuristic is upper bounded by the minimum of HSO and BR.

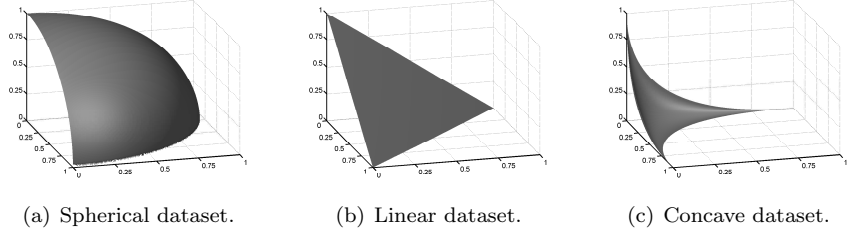

 (a) Spherical dataset.    (b) Linear dataset.    (c) Concave dataset.

Figure 1: Visualization of the first three datasets.

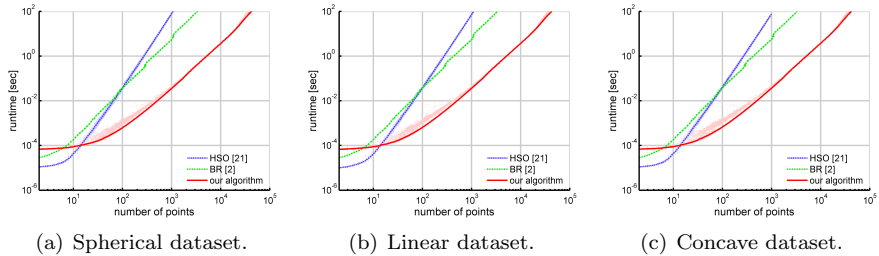

 (a) Spherical dataset.    (b) Linear dataset.    (c) Concave dataset.

Figure 2: Experimental results for $d = 3$.

Note that this improvement changes nothing for high dimensions (say, $d > 20$) as both exact algorithms quickly become unusable for these cases. The observed power of our algorithm for high dimensions (like $d = 100$) comes from the sampling, not from the combination with the exact algorithms.

## 4 Experimental analysis

To demonstrate the performance of the described approximation algorithm for the hypervolume contribution, we have implemented it and measured its performance on different datasets. We now first describe the used benchmark datasets and then our results.

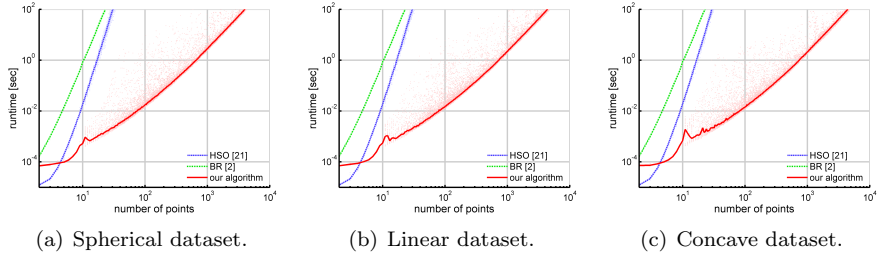

Figure 3: Experimental results for $d = 10$.

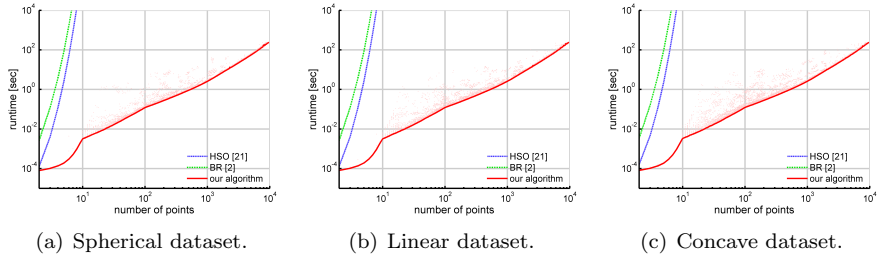

Figure 4: Experimental results for $d = 100$.

## 4.1 Datasets

We used five different fronts similar to the DTLZ test suite [10]. As we do not want to compare the hypervolume algorithms for point distributions specific to different optimizers like NSGA-II [9] or SPEA2 [27], we have sampled the points from different surfaces randomly. This allows full scalability of the datasets in the number of points and the number of dimensions.

To define the datasets, we use random variables with two different distributions. Simple uniformly distributed random variables are provided by the build-in random number generator `rand( )` of C++. To get random variables with a Gaussian distribution, we used the polar form of the Box-Muller transformation as described in [18].

### 4.1.1 Linear dataset:

The first dataset consists of points $(x_1, x_2, \ldots, x_d) \in [0,1]^d$ with $\sum_{i=1}^{d} x_i = 1$. They are obtained by generating $d$ Gaussian random variables $y_1$, $y_2$, ..., $y_d$ and then using the normalized points

$$(x_1, x_2, \ldots, x_n) := \frac{(|y_1|, |y_2|, \ldots, |y_n|)}{|y_1| + |y_2| + \ldots + |y_d|}.$$

#### 4.1.2 Spherical dataset:

To obtain uniformly distributed points $(x_1, x_2, \ldots, x_d) \in [0,1]^d$ with $\sum_{i=1}^{d} x_i^2 = 1$ we follow the method of Muller [14]. That is, we generate $d$ Gaussian random variables $y_1, y_2, \ldots, y_d$ and take the points

$$(x_1, x_2, \ldots, x_n) := \frac{(|y_1|, |y_2|, \ldots, |y_n|)}{\sqrt{y_1^2 + y_2^2 + \ldots + y_d^2}}.$$

#### 4.1.3 Concave dataset:

Analogously to the spherical dataset we choose points $(x_1, x_2, \ldots, x_d) \in [0,1]^d$ with $\sum_{i=1}^{d} \sqrt{x_i} = 1$. For this, we generate again $d$ Gaussian random variables $y_1, y_2, \ldots, y_d$ and use the points

$$(x_1, x_2, \ldots, x_n) := \frac{(|y_1|, |y_2|, \ldots, |y_n|)}{(\sqrt{|y_1|} + \sqrt{|y_2|} + \ldots + \sqrt{|y_d|})^2}.$$

For $d = 3$, the surface of the dataset is shown in Figure 1. Additionally to random points lying on a lower-dimensional surface, we have also examined the following two datasets with points sampled from the actual space similar to the random dataset examined by While et al. [23].

#### 4.1.4 Random dataset 1:

We first draw $n$ uniformly distributed points from $[0,1]^d$ and then replace all dominated points by new random points until we have a set of $n$ nondominated points.

#### 4.1.5 Random dataset 2:

Very similar to the previous dataset, we choose random points until there are no dominated points. The only difference is that this time the points are not drawn uniformly, but Gaussian distributed in $\mathbb{R}^d_{\geq 0}$ with mean 1.

Note that the last two datasets are far from being uniformly distributed. The points of the first set all have at least one coordinate very close to 1 while the points of the second set all have at least one coordinate which is significantly above the mean value. This makes their computation for many points (e.g., $n \geq 100$) in small dimensions (e.g., $d \leq 5$) computationally very expensive as it becomes more and more unlikely to sample a nondominated point.

### 4.2 Comparison

We have implemented our algorithm in C++ and compared it with the available implementations of HSO by Eckart Zitzler [25] and BR by Nicola Beume [2, 3]. For this we ran both algorithms on the whole front once and for every point on the front once without that point, to calculate all contributions as the differences. We did not add any further heuristics to both exact algorithms as all published heuristics do not improve the *asymptotic* runtime and even a speedup of a few magnitudes does not change the picture significantly.

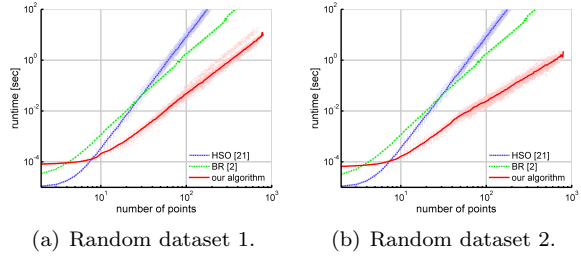

(a) Random dataset 1. (b) Random dataset 2.

Figure 5: Experimental results for random datasets with $d = 5$.

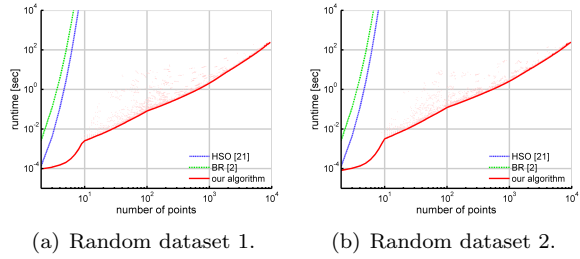

(a) Random dataset 1. (b) Random dataset 2.

Figure 6: Experimental results for random datasets with $d = 100$.

All experiments were run on a cluster of 100 machines, each with two 2.4 GHz AMD Opteron processors, operating in 32-bit mode, running Linux. For our approximation algorithm we used the parameters $\delta = 10^{-6}$ and $\varepsilon = 10^{-2}$. The code used is available upon request and will be distributed from the homepage of the second author.

Figure 2-6 show double-logarithmic plots of the runtime for different datasets and number of dimensions. The shown values are the median of 100 runs each. To illustrate the occurring deviations below and above the median, we also plotted all measured runtimes as lighter single points in the background. As both axes are scaled logarithmically, also the examined problem sizes are distributed logarithmically. That is, we only calculated Pareto sets of size $n$ if $n \in \{\lfloor \exp(k/100) \rfloor \mid k \in \mathbb{N}\}$. We examined dimensions $d = 3, 10, 100$ for the first three datasets and $d = 5, 100$ for the last two datasets.

Independent of the number of solutions and dimensions, we always observed that, unless $n \leqslant 10$, our algorithm outperformed HSO and BR substantially. On the used machines this means that only if the calculation time was insignificant (say, below $10^{-4}$ seconds), the exact algorithm could compete. On the other hand, the *much* lower median of our algorithm also comes with a much higher empirical standard deviation and interquartile range. In fact, we observed that the upper quartile can be up to five times slower than the median (for the especially degenerated random dataset 1). The highest ratio observed between the maximum runtime and the average runtime is 66 (again for the random dataset 1). This behavior is represented in the plots by the spread of lighter datapoints in the back of the median. However, there are not too many outliers and even their runtime outperforms HSO and BR. The non-monotonicity of our algorithm around $n = 10$ for $d = 10$ is caused by the approximations for the runtimes of the exact algorithms.

For larger dimensions the advantage of our approximation algorithm becomes tremendous. For $d = 100$ we observed that within 100 seconds our algorithm could solve all problems with less than 6000 solutions while HSO an BR could not solve any problem for a population of 6 solutions in the same time. For example for 7 solutions on the 100-dimensional linear front, HSO needed 13 minutes, BR 7 hours while our algorithm terminated within 0.5 milliseconds.

Parallel to our work, Bader and Zitzler [1] presented an approximation algorithm for the hypervolume contribution based on user-defined confidence levels. To use it, one has to choose four parameters (fitness parameter $k$, maximum number of sampling points $M_{\max}$, desired confidence $L$, and sampling interval $\Theta$). Though in their experiments they always use a fixed number of sampling points, we still expect that there is a mapping from our two parameters $\varepsilon$ and $\delta$ to their set of parameters such that both algorithm eventually behave alike. However, we did not try to prove that such a mapping exists as we consider our algorithmic framework to be much simpler since the user only has to choose two parameters $\varepsilon$ and $\delta$ such that the probabilistic performance guarantee of equation (1) from page 7 matches his needs.

On the other hand, there recently also appeared the first exact algorithm for the hypervolume contribution [5]. Bradstreet et al. [5] examined it on random fronts and fronts from the DTLZ test suite [10] with $n < 1000$ and $d \leqslant 13$. The maximum speedup they observed for any such front compared to HSO was 50. This compares to speedups of our algorithm compared to HSO of more than 1000 already for fronts with $n = 20$ and $d = 10$. For more points or more dimensions we could not calculate the speedup factor as HSO becomes much too slow.

## 5  Conclusions

We have proven that most natural questions about the hypervolume contribution which are relevant for evolutionary multi-objective optimizers are not only computationally hard to decide, but also hard to approximate. On the other hand, we have presented a new approximation algorithm which works extremely fast for all tested practical instances. It can solve efficiently large high-dimensional instances ($d \geqslant 10$, $n \geqslant 100$) which are intractable for all previous exact algorithms and heuristics.

It would be very interesting to compare the algorithms on further datasets. We believe that only when two solutions have contributions of very close value, our algorithm slows down. For practical instances this should not matter as it simply occurs too rarely – but this conjecture should be substantiated by some broader experimental study in the future.

REFERENCES                                                              22